
\hsize26pc
\vsize36pc
\magnification=1200
\bigskip
\centerline{\bf Quantum Resistive Transition }
\centerline{\bf in type II Superconductors under
	     Magnetic Field}
\bigskip
\centerline{\rm Ryusuke Ikeda}
\smallskip
\centerline{\rm Department of Physics, Kyoto University, Kyoto 606-01}
\bigskip
\bigskip
\bigskip
\bigskip
\bigskip

 It is shown that, within a Ginzburg-Landau (GL) formalism,  the
 superconducting fluctuation is insulating at zero temperature
 even if the fluctuation
 dynamics is metallic (dissipative).   Based on this result,
 the low temperature behavior of the $H_{c2}$-line and the resistivity
 curves near a zero temperature transition are discussed.  In particular,
 it is pointed out that the neglect of quantum fluctuations in data
 analysis of the dc resistivity may lead to an
 underestimation of the $H_{c2}$ values near zero temperature.

\vfill\eject

  It is well understood$^{1}$ now that thermal fluctuations of the
  superconducting
 order parameter $\psi$ in a type II superconductor destroy the
 second order transition at the mean field $H_{c2}$ line and
 describe the resistive broadenings in the resulting vortex liquid (VL)
 regime below the $H_{c2}$ line.
  Since the $H_{c2}$ line,  more or less, approaches
 zero temperature ( $\!T=\!0$ ) and, at such high fields,  the fluctuations
 in the VL regime
 will be safely described  within
 the lowest Landau level(LL) where
 the phase of $\psi$ {\it cannot} be separated from
 its amplitude,  one needs to take account of {\it
 quantum} fluctuations
 of $\psi$  within a Ginzburg Landau (GL) model appropriate in such low
 temperatures and high fields.  To our best knowledge,  quantum fluctuation
 effects on the dc linear transport have not been studied in such a high field
 GL region.  At present, there are many experimental situations$^{2,3}$
 closely related to this theoretical subject.

    In this letter, we point out that the GL superconducting fluctuation in a
 disordered phase at $T=0$
 is {\it insulating}, even when its dynamics is metallic (dissipative),  and
 possibly, for arbitrary dynamics.  Namely, irrespective of
  characters of the quantum fluctuations in {\it static} phenomena,
  the dc diagonal conductivity
 $\sigma_{xx}$ in a $T\!=\!0$ disordered phase does not have any fluctuation
 enhancement $\sigma_{s,xx}$  even
 below the $H_{c2}$-line.  This is in contrast to the case with
 purely thermal fluctuations where fluctuation effects on static quantities
 are
 consistent with those on dc linear dissipations.
  Therefore, in the (quantum) VL
 regime at nonzero temperatures where the quantum fluctuations significantly
 mask the reduction of the thermal fluctuation accompanying cooling, the dc
 diagonal resistivity below the $H_{c2}$-line has to (again) show a broadening
 which becomes more remarkable upon cooling. Consequently, in the VL regime at
 such low temperatures, the simple vortex dynamics
 based on a force balance$^{4}$
 is not applicable even qualitatively.

  First, we present the derivation of the $T=0$ result
  ${\sigma_{s,xx}}\!=\!{\sigma_{xx}}\!-\!{\sigma_{n,xx}}$
  (the normal conductivity) $\!=\!0$
  , and its relevance to real experiments
  will be discussed at the end of the paper.  As is seen below, the analysis
  leading to this result is, in a sense, transparent mathematically.  For
  simplicity,  we consider the usual time-dependent GL equation
  of $\psi$ (in the dimensionless unit) even at $T\!=\!0$
  and in strong fields;
  $(\gamma+{\rm i}\gamma')\partial\psi/{\partial t}=-[\varepsilon\psi+
  (-{\rm i}\partial-A)^2\psi+g{|\psi|^2}\psi]$(+ noise terms),
  where the constants $\gamma$  and $g$ are positive,
  and $A$ is the gauge field.
  Alternatively, we can work in the following Eucledean action$^{5}$
  (with $\hbar=1$)
  $$S={\int_r}\biggl[{\beta^{-1}}{\sum_\omega}(\gamma|\omega|
  +{\rm i}{\gamma'}\omega){|{\tilde \psi}_\omega|^2}+{\int_0^\beta}d\tau\,
  \biggl(\varepsilon{|\psi|^2}+{|(-{\rm i}\partial-A)\psi|^2}+{g\over 2}
  {|\psi|^4}\biggr)\biggr], \eqno(1)$$
  with $\psi(\tau)={\beta^{-1}}{\sum_\omega}{{\tilde \psi}_\omega}{\rm e}^{-
  {\rm i}\omega\tau}$ and $Z={{\rm Tr}_\psi}\exp(-S)$.
  Here $\omega$ is Matsubara
  frequency, and $\tau$ the imaginary time. Extending our proof
  to more realistic models is straightforward and,  together with quantum
  fluctuation effects on statics and detailed calculations relevant to
  resistivity data, will be given elsewhere$^{6}$.
  Below, fluctuation contributions to be absorbed into the normal conductivity
  and unrelated to the GL description are assumed to be negligible,
  because they have
  nothing to do with the vortex flow behavior$^{1}$ existing
  away from the quantum
  VL regime and hence, should not be essential to the ordering even at $T=0$.
  Further, we assume, as usual,  that a resummation scheme of Feynman
  graphs, such as the perturbative renormalization
  group,  can accurately describe fluctuation effects
   on the linear responses.  Then,  the dissipative $\psi$-dynamics
   is indispensible$^{6}$  to obtaining nonzero superconducting contributions
   to conductivities existing at nonzero
  temperatures. Hence, one needs knowledges on the dissipative
  quantum fluctuations at $T=0$  when focusing on nonzero temperatures at
  which a localization effect$^{7}$  on the $\psi$-dynamics
  may still be negligible.

  At nonzero fields,
  the order parameter field $\psi$ is Landau-quantized:
  ${{\tilde \psi}_\omega}\!=\!\!{\sum_{n, K}}{\phi_{n,K}}(\omega)$
  ${u_{n,K}}(r)$,
  where $n$ is the LL index, and other quantum numbers are specified by $K$,
  and $u_{n,K}$ is the LL eigen-basis.  Correspondingly, we have the
  renormalized fluctuation propagators
  ${{\cal G}_{n,K}}
  (\omega)=\langle|{\phi_{n,K}}(\omega)|^2\rangle$. Hereafter,
  the index $K$ playing no roles in the following discussions need not be
  denoted below.  We can assume the following form of ${\cal G}_n$ in the
  VL regime at $T=0$
  $${{\cal G}_n}(\omega)={1\over {\gamma|\omega|+{\rm i}{\gamma'_n}\omega
  +{d_n}}}={\int_{-\infty}^{\infty}}{dx}\,{{{\rho_n}(x)}\over
  {x-{\rm i}\gamma\omega}}, \eqno(2)$$
  where$^{5}$  ${\rho_n}(x)={{|a_n|^2}x}/(\pi{|x-{\rm i}{a_n}{d_n}|^2})$
  with ${a_n}=1/(1-{\rm i}{\gamma'_n}/{\gamma})$.
  By using this spectral representation (following the second equality)
   in each order of the perturbation series and noting that ${\rho_n}(x)$ is
  real, one can verify that there have
  no self energy
  corrections to the dissipative
  term $\sim|\omega|$ in (1).  Hereafter we set $\gamma=1$ below.
   The mass term in the static part $d_n^{-1}$ of
  ${\cal G}_n$ is always renormalized to a positive value in the VL
  regime,
   and the coefficient of the
  `Hall' term $\sim{\rm i}\omega$ is also
  renormalized and generally becomes $n$-dependent. For simplicity, we neglect
  other self energy corrections leading to higher order terms in $\omega$
  which, one can recognize$^{6}$,  does not change the result of this work.
  Due to the vanishing of the superfluid rigidity at zero wavevector in any
  direction, the superconducting (fluctuation) part $\sigma_{s,ij}$ of dc
  conductivity tensor can be found from the Kubo formula. In the case with a
  current ($\parallel x$) perpendicular to the magnetic field ($\parallel z$),
   its diagonal ($\sigma_{s,xx}$) and Hall ($\sigma_{s,xy}$)
   components are generally given by
  $$P(0)-P(\Omega)= |\Omega|\,{\sigma_{s,xx}}+\,{\rm i}\Omega
  \,{\sigma_{s,xy}}+O(\Omega^2), \eqno(3)$$
  where $P(\Omega)=T{\sum}{\sqrt{(n\!+\!1)(n'\!+\!1)}}
  \langle{\phi_{n+1}^*}(\omega\!+\!\Omega){\phi_n}(\omega)
  {\phi^*_{n'}}(\omega')
  {\phi_{n'+1}}(\omega'\!+\!\Omega)\rangle$
  is a polarization function (We do
  not need the details of the coefficient of $P(\Omega)$ below ),  and
  the general property ${{\cal G}_n}(-\omega)={{\cal G}_n^*}(\omega)$ was used.
    Hereafter, we can assume $\Omega>0$, and,  we
  only have to see the O($\Omega$) term of the real part of $P(\Omega)$,
  denoted by
  ${\partial_{\Omega}}{\rm Re}P|_0$ below,
   in order to find $\sigma_{s,xx}$.  First, let us calculate
  explicitly the
  Hartree term $P_{\rm H}(\Omega)=
  T\sum(n\!+\!1){{\cal G}_n}(\omega){{\cal G}_{n+1}}
  (\omega\!+\!\Omega)$ (i.e., the Aslamasov-Larkin graph)
  with no vertex corrections.  Using
  the spectral representation of (2),  one obtains  for arbitrary $T$
  $${P_{\rm H}}(\Omega)=\!{\sum_n}(n+1){f_{{\rm H},n}}(\Omega)=\!{\sum_n}(n+1)
  \,\,T{\sum_\omega}{{\,{p_{n,\,n+1}}(\omega,\Omega)+
  {p^*_{n+1,\,n}}(\omega, \Omega)\,}\over 2}, \eqno(4)$$
  where $${p_{n,\,n'}}(\omega,\Omega)=\!{{a_n}\over {|\omega|\!+\!{a_n}{d_n}
  \!+\!\Omega}}\biggl(
  {{a_{n'}}\over {|\omega|\!+\!{a_{n'}}{d_{n'}}}}+{{{a^*_{n'}}\,\Omega}\over
  {({a_n}{d_n}\!+\!{a^*_{n'}}{d_{n'}}\!+\!\Omega)(|\omega|\!+\!{a_n}{d_n})}}
  \biggr). $$
  The zero field case of (4) was derived previously by Tsuzuki$^{8}$ focusing
  only
  on nonzero temperatures.
  The $\omega\!=\!0$ term of ${\partial_{\Omega}}{\rm Re}{P_{\rm H}}|_0$
  directly
  gives the Hartree result in the thermal case,  and, deep in the thermal
  VL regime,  their real and imaginary
  parts can give,
   respectively, the diagonal and Hall vortex flow
  conductivities.  Hereafter we focus on $T\!=\!0$.
    One can verify from (4) that
  the ${\partial_{\Omega}}{\rm Re}{P_{\rm H}}|_0$  at $T\!=\!0$
  is precisely zero,  implying ${\sigma_{s,xx}}\!=\!0$ up to the Hartree
  level.
  The corresponding imaginary part of (4) (i.e., $\sigma_{s,xy}$ at $T=0$),
  which is nonvanishing in general,
  will be discussed elsewhere$^{6}$. We note that higher order terms in
  $|\Omega|$ (such as O($|\Omega|^3$) one) are nonvanishing in the real part
  of (4). To understand this, we rewrite (4) into another form following from
  the $\omega$-integral
  $${\rm Re}{f_{{\rm H},n}}(\Omega)=-{\int_0^\infty}dx
  {\int_0^\infty}dy\,({{\rho_n}(x)}{{\rho_{n+1}}(-y)}+
  {{\rho_n}(-x)}{{\rho_{n+1}}(y)}){{x+y}\over {{(x+y)^2}+{\Omega^2}}}.
  \eqno(5)$$
  Due to the power counting of the parameters $x\!$ and $\!y$,
  one finds that,
  if (5) is expanded in powers of $\Omega^2$, its O($\Omega^{2p}$) terms
  have divergent integrals for vanishing $x$ and $y$ if $p\geq 2$, implying
  nonanalytic appearances of $|\Omega|^{2p+1}$ ($p\geq 1$) terms in (5).
 Namely, only the ${\partial_{\Omega}}{\rm Re}{P_{\rm H}}|_0$, in a sense,
 trivially vanishes.

  Actually, the vanishing of ${\partial_{\Omega}}{\rm Re}P|_0$ (i.e.,
  of $\sigma_{s,xx}$) can be seen more simply in {\it arbitrary} graph
  of $P(\Omega)$ in the
  following way$^{6}$.   First, let us apply the identity
  ${\delta_{\sum{\omega_j}, 0}}=
  {\int_s}\exp({\rm i}{\sum{\omega_j}}s)$ to the frequency conservation at all
  bare vertices (including the current vertices) appearing in each graph,
  and divide ${{\cal G}_n}(\omega)$
  into its real and imaginary parts, where the real (imaginary) part is even
  (odd) with respect to $\omega$ due to
  the property
  ${{\cal G}_n}(-\omega)={{\cal G}_n^*}(\omega)$.
  Then, taking account of the fact that
  arbitrary Feynman graph of $P(\Omega)$ includes even number
  of ${\cal G}$'s,  it is
  easy to find that, after performing all frequency-integrals in each
  graph, the O($\Omega^{2p+1}$) terms of the resulting
  expression becomes pure
  imaginary.  However, it is not clear
  from this discussion whether a
   nonanalytic origin, in the sense of (5),  of
   ${\partial_{\Omega}}{\rm Re}P|_0$ is also impossible
  or not. In order to examine this, let us apply similar discussions to that
  for (5) to lower order graphs in perturbation series.  For instance, the
  corresponding expression to (5) resulting from the lowest (second)
   order vertex corrections becomes
   $$\delta{{\rm Re}{\bar P}_2}(\Omega)\sim \int{\prod_{i=1}^4}{dx_i}\,{x_i}\,
   {{{\sum_{j=1}^4}x_j}\over {{({\sum_{j=1}^4}x_j)^2}+{\Omega^2}}}, \eqno(6)$$
  where the contributions leading to (if any) nonanalytic O($|\Omega|^{2p+1}$)
  terms were
  merely considered here,  and the property at small $x$ of the spectral
  function, ${\rho_n}(x)\sim x$, characteristic of
   the Ohmic dissipation in $\psi$-dynamics, was used.   Due to the
  fact that (6) does not have any divergence
  for vanishing $x_j$'s  up to O($\Omega^4$),
  even an O($|\Omega|^3$) term does not follow from (6). Since this
  discussion is based on the power
  counting and hence systematic,  other higher order terms of
  the vertex correction
  cannot bring any nonanalytic appearance of
  ${\partial_\Omega}{\rm Re}P|_0$.   Therefore, ${\sigma_{s,xx}}=0$
  in general in a $T=0$ disordered phase.

   This result is essentially independent of the size of
   the field and trivially holds for the conductivity
   $\sigma_{zz}$ parallel to the field.
   Further, a similar analysis can also
   be used for fluctuation corrections to the mean field vortex flow
   conductivity $\sigma_{\rm MF}$ in the vortex {\it lattice}
   at $T\!=\!0$ with
   phase long ranged order and hence,
   $\sigma_{s,xx}$ at $T\!=\!0$ jumps discontinuously
   at the melting transition
   field ${H_m}(0)$ from $\sigma_{\rm MF}$ to zero
   ( in the disordered case$^{7}$,
   from $\infty$ to zero ). The resulting picture on the resistive transition
   in 2D case near $T=0$
    is sketched in Fig.1,
   where it is, for simplicity, assumed that the (extrapolated) normal
   resistance $R_n$
   and $H_{c2}$ are temperature-independent and that the
   Hall coefficient is negligibly small.  Reflecting the
   above $T=0$ result on $\sigma_{s,xx}$,
    the resistivity above ${H_m}(0)$ may
   show a quasi-reentrant behavior at nonzero temperatures, of which the
   presence has been often speculated$^{9}$ within a
   phase-only model for granular systems in $H=0$.  However, an insulating
   behavior of
   $\sigma_{n,xx}$ may interrupt observations of such a behavior
   of $\sigma_{xx}$ in experiments.

   The present result on superconducting fluctuations tends to {\it support}
    the suggestions$^{2}$,  based on resistivity data, of an upwardly
    curved $H_{c2}$-line
    at low $T$:
    It is not difficult to imagine that, in general,  such an
   $H_{c2}$-determination
    in the temperature range with significant quantum fluctuations
   tends to result in an (incorrectly) underestimated $H_{c2}$-value
   because,  according to our result,  a vestige of
   the (correct) $H_{c2}$-line should gradually dissappear from
   dc resistive data as $T\to 0$.  A rapid variation of the
   dc resistivity in the quantum regime should, in the clean limit,
   occur near the true
   (melting) transition,  while the $H_{c2}$-line should be reflected,
   even when $T\to 0$,  in the magnetization$^{6}$.   Such an increase of
   $H_{c2}$ near $T=0$ becomes  another origin of covering the
   quasi-reentrant behavior in experimental data.
   Nevertheless,
   the data in Ref.2 at lowest
   temperatures have shown broadenings becoming remarkable upon cooling,
   which can be
   explained$^{6}$ based on the present result.  We note that a
   broadening in real systems with randomness at low $T$ cannot be explained
   without quantum superconducting fluctuations,  which provide interactions
   among the order parameter fields in the quantum VL regime.  Any approach$^
   {10}$ of such phenomena based on the mean field theory is theoretically
   invalid.  The strongly disordered case$^{3}$  will be
   discussed elsewhere$^{6}$.
   We merely note,  together with the importance of dissipation mentioned
   in the introduction,  that  there are no reasons why the scenario based on
   a phase-only model$^{7}$  is applicable in strong field regime where
   the amplitude fluctuation of the order parameter is {\it qualitatively}
   quite different from that in low field regime.

\bigskip

   The author is grateful to Y. Ando for useful conversations and to Steve
   Girvin and Department
   of Physics, Indiana University for hospitality where this manuscript was
   written.  This work
   was finantially supported by the Grant-in-Aid for Scientific Research from
   the Ministry of Education, Culture, and Science in Japan.

\vfill\eject

\leftline{References}
\frenchspacing
\bigskip
\item{1} R.~Ikeda, J.Phys.Soc.Jpn. {\bf 64}, 1683(1995).
\item{2} For instance, M.S.~Osofsky et al., Phys.Rev.Lett. {\bf 71},2315(1993)
\item{}  ; R.~Yoshizaki et al., Physica {\bf C224}, 121(1994).
\item{3} For instance, A.F.~Hebard and M.A.~Paalanen, Phys.Rev.Lett. {\bf 65},
\item{}  927(1990).
\item{4} G.~Blatter et al., Rev.Mod.Phys. {\bf 66}, 1125(1994).
\item{5} E.~Abrahams and T.~Tsuneto, Phys.Rev.B{\bf 11}, 4498(1975).
\item{6} R.~Ikeda, in preparation.
\item{7} M.P.A.~Fisher, Phys.Rev.Lett.{\bf 65}, 923(1990).
\item{8} T.~Tsuzuki, Prog.Theor.Phys. {\bf 42}, 1020(1969).
\item{9} For instance, M.P.A.~Fisher, Phys.Rev.B{\bf 36}, 1917(1987).
\item{10} For instance, G.~Zwicknagl and J.W.~Wilkins, Phys.Rev.Lett.
\item{}  {\bf 53}, 1276(1984).

\vfill\eject

\leftline{Figure Caption}
\frenchspacing

\item{Fig.1}  Schematic temperature variations of 2D resistance curves at low
$T$ and at each
 field $H$,  drawn by assuming
the normal resiatance $R_n$ and the $H_{c2}$
line to be $T$-independent. The dotted (solid) curves are
for ${H_{c2}}(0)\geq H>{H_m}(0)$ ($H<{H_m}(0)$).

\bye